\documentclass{aa}

\usepackage{lineno}

\usepackage{graphicx}

\usepackage{txfonts}

\usepackage[table,xcdraw]{xcolor}
\usepackage{hyperref}
\hypersetup{colorlinks=true, citecolor=blue}
\usepackage{booktabs}

\defcitealias{tissera2022}{T22}

\newcommand{\mdotbh}{\dot{M}_{\rm BH}}

\newcommand{\mbh}{M_{\rm BH}}
 
\newcommand{\mdotedd}{\dot{M}_{\rm Edd}}

\newcommand{\Ledd}{\lambda_{\rm Edd}}

\newcommand{\cMpc}{\rm cMpc}

\newcommand{\lsim}{\mathrel{\hbox{\rlap{\lower.55ex\hbox{$\sim$}} \kern-.3em\raise.4ex\hbox{$<$}}}}
\newcommand{\gsim}{\mathrel{\hbox{\rlap{\lower.55ex\hbox{$\sim$}} \kern-.3em\raise.4ex\hbox{$>$}}}}

\newcommand{\REF}{Ref-L100N1504}

\newcommand{\RECAL}{Recal-L025N0752}

\newcommand{\eagle}{{\sc eagle}}
\newcommand{\dexkpc}{dex~kpc$^{-1}$}
\newcommand{\dexReff}{dex~$R_{\rm eff}^{-1}$}

\newcommand{\taustar}{$\tau_{25}$}
\newcommand{\Reff}{$R_{\rm eff}$}
\newcommand{\Ropt}{$R_{\rm opt}$}

\newcommand{\gradYS}{$\nabla^{\rm YS}$}
\newcommand{\gradSF}{$\nabla^{\rm SFG}$}
\newcommand{\gradSFR}{$\nabla_{\rm SFR}$}

\newcommand{\subfind}{{\sc subfind}}

\newcommand{\mstar}{$M_{\star}$}
\newcommand{\Msun}{$\rm M_{\sun}$}
\newcommand{\oh }{$ 12 + \rm log (O/H)$}

\newcommand{\HII}{H\,\textsc{ii}}
\newcommand{\fgas}{$f_{\rm{gas}}$}
\newcommand{\dep}{$\tau_{\rm{dep}}$}

\usepackage[normalem]{ulem}

\begin{document}

    \title{Aligned and misaligned metallicity gradients in young stars and star-forming regions in the EAGLE discs}
    \titlerunning{Aligned and misaligned metallicity gradients in YSs and SFG}

    \author{Isha Shailesh 
          \inst{1}\fnmsep\inst{2}\thanks{E-mail: ishailesh@uc.cl}
          \and Patricia B. Tissera\inst{1}\fnmsep\inst{2} \and Emanuel Sillero \inst{1}\fnmsep\inst{2}
          }
    
    \institute{Instituto de Astrof\'isica, Pontificia Universidad Cat\'olica de Chile. Av. Vicu\~na Mackenna 4860, Santiago, Chile.
    \and
    Centro de Astro-Ingenier\'ia, Pontificia Universidad Cat\'olica de Chile. Av. Vicuña Mackenna 4860, Santiago, Chile.
    }

    \date{Received XXX; accepted YYY}

  \abstract
   {Disc galaxies exhibit radial metallicity gradients in both their stellar and gaseous components. The star-forming gas (SFG) in \HII~regions and young stars (YSs) trace the recent evolutionary history of the galaxy. By analysing their metallicity gradients in tandem, we can shed light on the different physical processes that interact in a complex manner, modifying the distribution of chemical elements in the discs, thereby influencing the chemical evolution of galaxies.}
   {We aim to assess the extent to which the joint analysis of metallicity gradient alignment in YSs and SFG can constrain the recent evolutionary history of galaxies.}
   {Using the high-resolution run of the \eagle~project, we derived radial, azimuthally averaged oxygen abundance profiles for YSs (age $<2$ Gyr) and SFG and measured their gradients as the slopes of linear fits to these profiles. We classified galaxies into four groups based on the signs (N for negative and P for positive) of the slopes: NN, NP, PP, and PN (the first letter is for YSs and the second for SFG).}
   {We found that galaxies with NN, NP, PP, and PN combinations of metallicity profiles reflect different evolutionary paths over the past $\sim2$ Gyr. NN galaxies exhibit sustained inside-out growth accompanied by high star formation efficiency, whereas NP and PP systems show evidence of recent or ongoing feedback-driven disruption, with PP galaxies likely being predominantly shaped by supernova feedback. PN galaxies, by contrast, show evidence of past violent events followed by gradient recovery, highlighting the interplay between inflows, feedback, and gas cooling in shaping metallicity distributions.}
   {The degree of alignment between the stellar and gas metallicity gradients provides a way to time the occurrence of significant events in the evolutionary history of galaxies, which contribute through a combination of gas inflows, star formation triggering, and metal mixing. They could also serve as probes of sub-grid physics when observations provide suitable comparison datasets.}


   \maketitle



\section{Introduction}

In the Lambda cold dark matter ($\Lambda$CDM) paradigm, galaxies form within hierarchically assembled dark matter halos in an expanding universe \citep{white1978, white1991}. The evolution of galaxies is driven by a complex interplay of physical processes, acting on different temporal and spatial scales. The chemical composition of the interstellar medium (ISM) and the stellar populations in galaxies could store information on these processes \citep{tinsley1980, matteucci1986}. Chemical elements are manufactured in stellar interiors and expelled into the ISM via stellar winds, planetary nebulae, and supernova (SN) explosions, which enrich the ISM and are later incorporated into future generations of stars, continuing the cycle of chemical enrichment. 

The distribution of chemical elements in a galaxy is usually quantified by its metallicity profile\footnote{Metallicity is commonly used in astronomy to refer to the abundance of chemical elements heavier than He and is defined as the total mass of these elements in the gas or stellar component relative to the gas mass or stellar mass. We use the terms metallicity gradient and chemical abundance gradient interchangeably.} \citep[e.g.][]{garnett1997}. Negative metallicity gradients in galaxies, wherein the centre of a galaxy is more metal-rich than the outskirts, are expected in an inside-out formation scenario as in a $\Lambda$CDM framework. The predominance of observed negative gas-phase metallicity gradients at $z\sim0$ \citep[e.g.][]{shaver1983, henryworthey1999} strongly supports a scenario in which star formation begins in the central regions and progressively moves outwards as the galaxy grows, allowing more time for the accumulation of chemical elements in the core \citep[e.g.][]{sanchez2014, belfiore2017}. However, these gradients are also modulated by gas infall, outflows, and galaxy interactions and mergers \citep{pilkington2012, ma2017grad, yates2021, tissera2022, sharda2023}. Hence, metallicity gradients could serve as tracers of the physical processes that drive the evolution of galaxies \citep{ho2015, sillero2017, hemler2021, tapia2025}. 

Observations provide estimates of the metallicity distribution in \HII~regions, which serve as tracers of current metallicity in star-forming regions from the local to very high-redshift Universe \citep[e.g.][]{vallini2024}. In contrast, the estimation of chemical abundances of stellar populations is more complex, as in most cases only the integrated chemical history of stellar populations can be retrieved \citep[e.g.][]{sanchezb2014, frasier2022, greener2022, campsfarina2022}. Until recently, individual stellar abundances could only be observed within the Milky Way (MW; see \citealt{molla2019} for a compilation of MW data). However, chemical measurements for some individual stars in nearby galaxies are now becoming available \citep{sextl2024}. Combining information from star-forming regions and young stellar populations provides a deeper insight into the timescales of metal mixing and the physical processes that regulate star formation. For example, using numerical simulations, \citet{lian2023} reported negative gradients for both the young stars (YSs) with ages $<4$ Gyr and the star-forming gas (SFG) phase for MW-like stellar mass galaxies in the TNG50 simulations. However, these types of studies are still in their early stages.

In undisturbed galaxies, \HII~regions determine negative metallicity gradients \citep{belfiore2017}. However, there is a fraction of reported galaxies that exhibit \HII~regions with inverted (i.e. positive) metallicity gradients (i.e. their central regions are less enriched than the outskirts; \citealt{belfiore2017}). These galaxies are more frequently observed at high redshifts, where systems are more gas-rich and their environments are more turbulent and complex \citep{troncoso2014}. Galaxies can have shallow or inverted SFG metallicity gradients due to a recent inflow of metal-poor gas into central regions or the presence of strong galactic outflows. Galaxy-galaxy interactions have been shown to be very efficient in driving gas inflows \citep{Barnes&Hernquist1996}. Pristine or low-metallicity gas has also been shown to contribute to diluting the central metallicity \citep{ceverino2016}. While inflows of low-metallicity gas could flatten metallicity gradients, the accumulation of gas flowing into the central region could induce star formation activity \citep{tissera2000, sillero2017, moreno2019}. Massive newborn stars would soon inject metals, increasing the level of enrichment in the central regions and making the negative gradients steeper \citep{sextl2024}. These newborn stars would also inject energy, which could mix or expel chemical elements via mass-loaded galactic outflows \citep{dekelsilk1986}, and flatten or even invert the metallicity profiles \citep{perez2011, torrey2012}. Additionally, the formation of bars could induce stellar migration, which could change the metallicity distribution of the stellar populations. However, these mechanisms act on longer timescales and are expected to have a larger effect on old stellar populations \citep{minchev2014, tissera2017,rosas-guevara2025,fragkoudi2025}. 

Although SN feedback has long been considered the main driver of chemical evolution \citep{pilkington2012}, it is well accepted that feedback from the active galactic nucleus (AGN) plays a role in regulating star formation activity in massive galaxies, but its action in lower mass systems remains unclear. Its effect on the distribution of chemical elements also remains to be understood. AGN feedback may indeed contribute to modulating this distribution or even to inverting metallicity gradients; however, its overall impact remains a topic of ongoing debate \citep{Taylor&Kobayashi2017, doNascimento2022, jara2024, Amiri2024}.

When metallicity gradients are flattened or inverted, the re-establishment of negative slopes in the SFG phase indicates the onset of renewed star formation activity across a galaxy. This activity is likely fuelled either by previously heated gas that has cooled and becomes available for star formation in the central regions \citep{tapia2025}, or by the accretion of low-metallicity gas in the outer regions \citep{collacchioni2019, palla2024}. Additionally, the rate of change in the metallicity gradients depends on other factors such as gas richness, merger impact parameters, and the relative contribution of gas inflows, all of which can modulate the evolution of the gradients. These mechanisms could also modify the distribution of the star formation activity in the discs. \citet{tissera2022}, hereafter \citetalias{tissera2022}, reported a trend for strong negative and positive SFG metallicity gradients to be associated with violent events such as significant mergers or strong gas inflows. They estimated an average recovery time of about $1.4$ Gyr, ranging approximately within  $0.7-2$ Gyr, for the strong metallicity gradients to become weak again. 

In this work, using the high-resolution run of the \eagle~simulations, we explored the metallicity gradients of YSs in relation to those of SFG, focusing on the extreme cases where these two gradients are aligned or misaligned. The case of alignment refers to the metallicity profiles of both YSs and SFG having the same direction or sign, whereas in the case of a misalignment, the metallicity profiles of YSs and SFG have opposite directions or signs. We speculate that if star formation had been efficient, the newly formed stars would trace the physical conditions of the SFG regions at the time of their birth. Galaxies that recover the SFG negative-slope metallicity profile rapidly might not be able to invert the profile of their stellar populations born within the last $2$ Gyr. This age threshold was adopted for YSs taking the maximum recovering time reported by \citetalias{tissera2022} as a reference. Aligned metallicity gradients between YSs and SFG may suggest an equilibrium stage among gas inflows, star formation, and outflows, and the lack of violent events in the last few gigayears of evolution at least \citep{Peebles&Shankar2011, Dayal2013, lilly2013, forbes2014}. Conversely, differences in the alignment of these two components could indicate specific stages of evolution and provide estimations for mixing timescales, offering a deeper understanding of the galactic chemical-enrichment history. Hence, in this study, we aim to jointly study the signal of metallicity gradients of YSs and SFG and comparatively explore the properties of galaxies with the same and different alignment signals in order to infer the information they can store about their recent evolutionary history. Additionally, comparing these two metallicity profiles offers valuable insight into the sub-grid physics implemented in simulations of galaxy formation, especially as increasingly detailed observational data for external galaxies will become available in the near future, opening the possibility for a more extended comparison with observations.

This paper is organised as follows. In Sect. \ref{sec:galaxy_catalogue} we briefly summarise the main characteristics of the \eagle~suite of cosmological simulations used. In Sect. \ref{sec:results} we present our results from investigating the dependence of the metallicity gradients on galaxy properties, recent violent events, and AGN feedback. The results of this paper are discussed and summarised in Sect. \ref{sec:conclusion}.

\section{The galaxy catalogue}
\label{sec:galaxy_catalogue}

In our analysis, we used the galaxy catalogue built by \citetalias{tissera2022} from the higher resolution simulation \RECAL~of the \eagle~Project \citep{schaye2015,schaller2015} and studied the metallicity gradients of YSs (age $<2$ Gyr) and SFG. As \eagle~discs show a range of metallicity gradient combinations in YSs and SFG, this could contribute to dating the impact of recent events that significantly alter the metallicity profiles by redistributing metals and changing the thermodynamic properties of the ISM. The \eagle~simulations, particularly the \RECAL~run, allowed us to study a statistically significant sample of global metallicities and metallicity gradients in a cosmological context, in agreement with observational estimates for galaxies at local redshift $(z \lesssim 0.2)$ with \mstar~$>10^9$ \Msun, as shown by previous works such as \citet{schaye2015}, \citet{derossi2017}, \citet{tissera2019, tissera2022}, and \citet{jara2024}.

\subsection{The \RECAL~simulation}
The \eagle~Project assumes a $\Lambda$CDM universe with $\rm \Omega_{\Lambda} = 0.693$, $\rm \Omega_{m} = 0.307$, $\rm \Omega_{b} = 0.04825$, $h = 0.6777$ ($H_{\rm 0} = 100\ h$ $\rm km\ s^{-1}\ Mpc^{-1}$), $\rm \sigma_{8} = 0.8288$, $n_{\rm s} = 0.9611$, and $Y = 0.248$\footnote{The database is publicly available; see \citet{mcAlpine2016}.}. The \RECAL~run  has a comoving box size of $25\,\cMpc$, and has $752^{3}$ initial dark matter and baryonic particles with a mass resolution of $1.21\times 10^{6}$ \Msun~and $2.26\times 10^{5}$ \Msun, respectively.

All \eagle~simulations were performed with the ANARCHY sub-grid models grafted onto a version of {\sc p-gadget-3} \citep[][see for more details on the sub-grid physics]{schaye2015,crain2015}. Briefly, the sub-grid model incorporates cooling, star formation, metal enrichment, and SN feedback \citep{schaye2008,schaye2010,wiersma2009a,dallavecchia_schaye2012}, as well as black hole (BH) growth and feedback \citep{rosas-guevara2015}. The adopted initial mass function adheres to \citet{chabrier2003}, with minimum and maximum mass thresholds at 0.1 \Msun~and 100 \Msun, respectively. The chemical enrichment model tracks eleven different chemical elements \citep{wiersma2009}. A detailed comparison of the performance of \eagle~with other codes can be found in \citet{schaller2015}. 

The BH sub-grid model in \eagle~estimates the accretion rate using a modified Bondi--Hoyle formalism  that includes angular momentum suppression and is capped at the Eddington limit \citep{rosas-guevara2015}. BHs can also grow through mergers with other BHs when they are gravitationally bound and in the vicinity of each other. The AGN feedback was implemented as single-mode thermal (stochastic) feedback following \citet{Booth2009}.

\subsection{The selected galaxies}
\label{subsec:selected_galaxies}

From the \citetalias{tissera2022} catalogue, we selected a sample of galaxies with $z=[0,0.18]$, \mstar~$=[10^{9},10^{11}]$ \Msun, and  $\mathrm{SFR} \approx [0.08,7.3]$ \Msun~$\mathrm{yr}^{-1}$. Galaxies were required to contain at least 500 stellar particles within 1.5 \Ropt~to ensure a well-defined angular momentum, where \Ropt~is defined as the radius that encloses approximately $80\%$ of the stellar mass initially identified by the \subfind~algorithm. This sample of galaxies encompasses only central galaxies (i.e. the most massive galaxies within the virial halo).

This catalogue provided us with the following parameters:

\begin{itemize}

\item \mstar: the total stellar mass within 1.5 \Ropt

\item \Reff: the effective radius (in kiloparsecs) estimated as the radius that encloses $50\%$ of the stellar mass within 1.5 \Ropt

\item $R_{200}$: the galactocentric radius within which the mean enclosed density is 200 times the critical density of the Universe
\item $M_{200}$: the total mass contained within $R_{200}$

\item $\rm{sSFR}= {SFR}/$\mstar: the specific star formation rate (sSFR), where the SFR is the star formation rate of a given galaxy

\item $f_{\rm{gas}}= M_{\rm{gas}}/(M_{\rm{gas}}+$\mstar): the gas fraction within 1.5 \Ropt

\item $\tau_{\rm{dep}} = M_{\rm{gas}}/\rm{SFR}$: the depletion time

\item D/T: the disc-to-total stellar mass ratio, which was estimated by adopting the AM-E method\footnote{In the AM-E method, the stellar (gas) discs are defined by star (gas) particles with $\epsilon = J_{\rm{z}}/J_{\rm{z/max}}(E)>0.5$, where $J_{\rm{z/max}}(E)$ is the maximum angular momentum along the main axis of rotation, $J_{\rm{z}}$, over all particles at a given binding energy of $E$.} \citep{tissera2012}

\item \taustar: the time elapsed since the last increase in stellar mass by more than 25\%

\item \gradSF~(in \dexkpc): the azimuthal-averaged metallicity gradient of the SFG obtained by using the oxygen abundance, \oh, within the radial range [0.5,1.5] \Reff. The chemical abundances were weighted by the SFR in order to mimic observations. The \gradSF~was estimated only for galaxies with more than 1000 SFG particles. These gradients were also used by \citet{jara2024} to study a secondary dependence of the mass-metallicity relation on the metallicity gradients. 

\end{itemize}

We followed the procedure used by  \citetalias{tissera2022} for the  calculations of \gradSF~to similarly estimate the metallicity gradients of stars younger than $2$ Gyr, \gradYS~(in \dexkpc), within the same radial range [0.5,1.5] \Reff. The profiles were built from median oxygen abundances of YSs in 20 radial bins and fitted with a linear regression.\footnote{Since we compared the metallicity gradients of young stellar populations (\gradYS) with those of the gas (\gradSF), we used the same metallicity indicator, $12+\mathrm{log_{10}}\mathrm{(O/H)}$, for both components to allow a direct comparison between them.} To fit a linear regression, we required the profiles to be determined by at least five radial bins, corresponding to a minimum of 100 young stellar particles. This gave us a final sample of 239 central star-forming galaxies. This procedure is consistent with the method used for estimating stellar metallicity gradients in \citet{solar2020}. The median metallicity gradients of the galaxy sample are given in Table~\ref{tab:table_stats_gradients}.

We note that it has been reported that galaxies in the \eagle~simulations tend to be deficient in bulge components when compared with observations \citep[e.g.][]{degraaff2022}. This issue may raise concerns regarding the relevance of analyses that depend on the internal structural properties of galaxies. However, the simulation analysed in \citet{degraaff2022}, \REF, differs from the higher resolution \RECAL~run employed in this work, which includes re-calibrated sub-grid physics that can affect the spatial distribution of baryonic matter. While a reduced bulge component may influence the determination of the D/T, particularly in low-mass galaxies, its impact on the metallicity gradients estimated here is expected to be limited. This is because gradients were measured over the radial range of $[0.5,\,1.5]\,R_{\mathrm{eff}}$, which generally excludes the central regions where bulge-associated stellar mass dominates.

We adopted a single power law to fit the metallicity gradients within a radial range of [0.5,1.5] \Reff~as is usually done, and to be consistent with \citetalias{tissera2022} and \citet{jara2024}, where this criterion was also adopted. However, a detailed description of the disc metallicity distribution might show the existence of inner and outer breaks \citep[e.g.][]{diaz1989, sanchezmenguiano2016, Garcia2024, tapia2025}. 

It is also important to note that past observations have reported azimuthal variations of the chemical abundances of the star-forming regions \citep{SancMen2019,Vogt2017}. Comparable azimuthal variations have also been detected in the \eagle~galaxies \citep{ScholzDiaz2021}. In particular, and related to our work, \citet{solar2020} analysed the azimuthal variations of metallicity gradients in the \eagle~simulations, finding a correlation between gradients measured along random directions and those derived from azimuthally averaged radial profiles of galactic discs (Spearman $r_{\mathrm{S}} \sim 0.46$). They reported a median azimuthal dispersion of $\sigma_{\nabla} \simeq 0.12 \pm 0.03~\mathrm{dex}\,R_{\mathrm{eff}}^{-1}$. These azimuthal variations do not significantly affect the azimuthally averaged radial profiles adopted in this work, as the averaging procedure smooths out local fluctuations. The persistence of the trends after averaging indicates the presence of a genuine underlying signal. We acknowledge that this signal may be asymmetric; however, a detailed analysis of the full two-dimensional metallicity structure would require a different methodological approach designed to fully exploit spatially-resolved information such as that provided by integral field spectroscopy observations \citep[e.g.][]{belfiore2017}. Such an analysis is beyond the scope of the present study.

\section{The statistics of metallicity gradients}
\label{sec:results}

\begin{figure}
    \includegraphics[width=0.9\columnwidth]{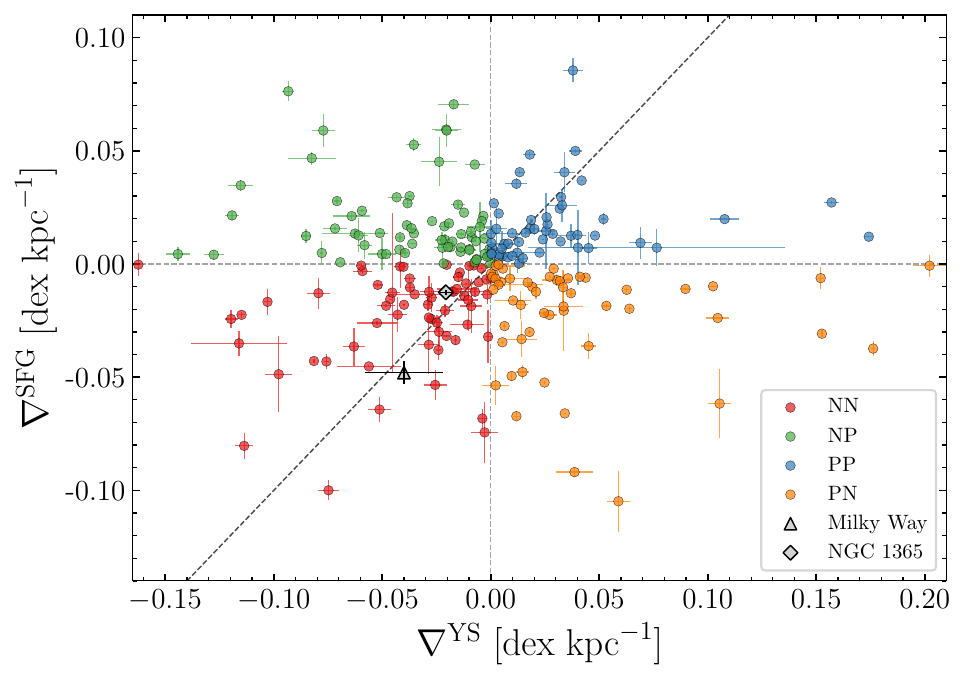}
    \caption{Metallicity gradients of SFG (\gradSF) as a function of metallicity gradients of YSs (\gradYS) in units of \dexkpc~selected from the \RECAL~\eagle~simulation. The error bars on the gradients were estimated from the standard deviation of the linear regressions. For comparison, the oxygen abundance gradients of YSs (age $\leq2$ Gyr) and \HII~regions for the MW \citep{molla2019} and NGC 1365 \citep{sextl2024} are displayed (black symbols). Both the MW and NGC 1365 lie close to the 1:1 relation (dashed black line). The \eagle~galaxies are colour-coded according to the quadrants to which they belong. This colour-coding scheme is followed in all subsequent plots.}
    \label{fig:quadrants}
\end{figure}

In Fig.~\ref{fig:quadrants} we display \gradSF~as a function of \gradYS; this does not show a clear correlation between the two. In fact, while there are some galaxies located close to the 1:1 relation, other galaxies show a diversity of behaviours. In most cases, the metallicity gradients are quite shallow. A lack of correlation is also found when the metallicity gradients are normalized by \Reff,~as can be seen in Fig.~\ref{fig:quadrantsnorm}. For comparison, we included the oxygen abundance gradients of YSs (i.e. age $\leq2$ Gyr) and of \HII~regions estimated for the MW \citep{molla2019} and for NGC 1365, a barred spiral galaxy of stellar mass \mstar$\sim10^{10.95}$ \Msun~\citep{sextl2024}, respectively.
 
We grouped our galaxy sample according to their \gradYS~and \gradSF~to classify them depending on whether the gradients are aligned (i.e. both gradients have the same sign or direction) or misaligned (i.e. they have opposite signs). This classification allowed us to examine their properties and highlight the differences between galaxies with different combinations. Since we are only concerned with the sign of the gradients, we divided the galaxies into four subsamples. In this notation, N denotes a negative gradient and P a positive gradient, with the first letter referring to the YSs and the second to the SFG, as follows:

\begin{itemize}
    \item NN galaxies: \gradYS~$<0$, \gradSF~$<0$    
    \item NP galaxies: \gradYS~$<0$, \gradSF~$>0$
    \item PP galaxies: \gradYS~$>0$, \gradSF~$>0$
    \item PN galaxies: \gradYS~$>0$, \gradSF~$<0.$
\end{itemize}

First, we examined the distribution of galaxies in each quadrant as a function of stellar mass. We note that the median value of \mstar~for each subsample is $\sim10^{9.75}$ \Msun. To account for the dependence of metallicity gradients on stellar mass \citep[e.g.][]{belfiore2017}, we divided the galaxies in each quadrant into two stellar mass bins, adopting $10^{9.75}$ \Msun~as the threshold. This stellar mass threshold was also used in \citet{jara2024}. For each subsample, galaxies were grouped in a low-mass interval of $[10^{9},10^{9.75}]$ \Msun~and a high-mass interval of $>10^{9.75}$ \Msun. We limited our analysis to these two mass bins due to the small number of galaxies in our sample.

\begin{figure*}
        \includegraphics[width=2\columnwidth]{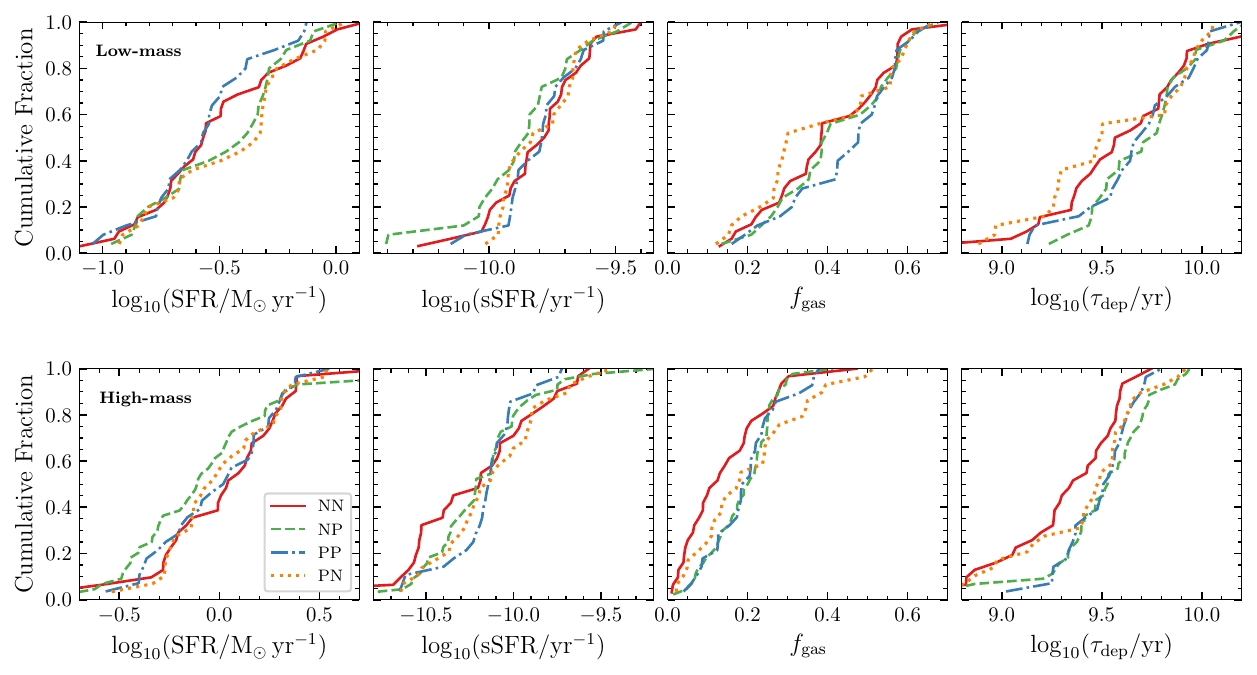}
    \caption{Cumulative distributions of the SFR, the sSFR, the gas fraction ($f_{\rm{gas}}$), and the depletion time ($\tau_{\rm{dep}}$) for low-mass (upper panel) and high-mass (lower panel) galaxies. Lines indicate galaxies classified as NN (solid red), NP (dashed green), PP (dash-dotted blue), and PN (dotted green).}
    \label{fig:cdf_all}
\end{figure*}

We analysed the relative distribution of galaxy properties in each subsample as a function of \mstar. In Fig.~\ref{fig:cdf_all} we show the cumulative fractions of SFR, sSFR, \fgas, and \dep. For low-mass galaxies, there is a weak trend for NN and PP galaxies to be less star-forming. There is a systematic trend for NN and PN galaxies to have slightly lower \fgas~and shorter \dep. They seem to be more actively and efficiently forming stars than the other two types of galaxy. High-mass galaxies have clearer trends. NN galaxies have higher SFRs and lower \fgas, which combine to produce shorter \dep. The rest of the galaxy types are more gas-rich and have lower SFRs, resulting in a slightly longer \dep.

\begin{figure}
        \includegraphics[width=0.9\columnwidth]{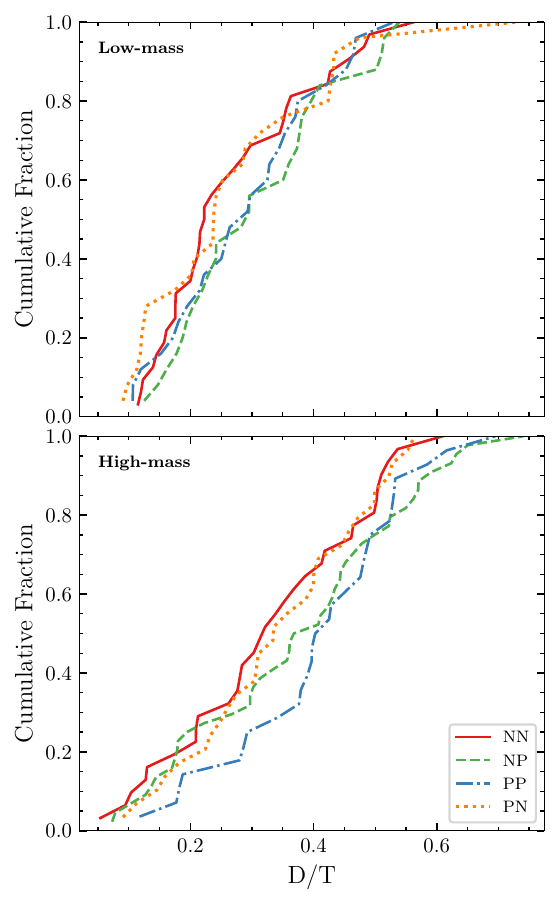}
    \caption{Cumulative distributions of disc-to-total stellar mass ratio (D/T) for low-mass (upper panel) and high-mass (lower panel) galaxies. Lines indicate galaxies classified as NN (solid red), NP (dashed green), PP (dash-dotted blue), and PN  (dotted orange).}
    \label{fig:cdf_DT}
\end{figure}

To complement this global analysis, in Fig.~\ref{fig:cdf_DT} we also show the cumulative fractions of the disc-to-total mass ratio (D/T) of galaxies for the two bins of stellar mass. NN and PN types have a lower fraction of disc-dominated systems with a more significant central bulge, regardless of the stellar mass. More centrally concentrated galaxies tend to have more massive bulges, which would make the disc more stable \citep{toomre1972}. Galaxies with inverted \gradSF~(i.e. NP and PP types) tend to have a higher fraction of more disc-dominated systems, which would provide them with a large gas reservoir, and they might be more unstable \citep[e.g.][]{toth1992,scan09,varelalavin2023}. However, NP and PP galaxies show larger \dep, implying that they consume their gas reservoirs in a steadier fashion than NN and PN galaxies. We also note that less massive systems are also more spheroidally dominated than higher mass systems.

In summary, among massive galaxies, we find that NN galaxies, followed by PN galaxies, exhibit higher SFRs, lower gas fractions, and shorter depletion times compared to other galaxy types. In contrast, PP and NP are more gas rich, but they seem to be less efficient at converting gas into stars. The trends present in low-mass systems are weaker.

\begin{table*}
\centering
\caption{Number of galaxies in each bin and the median, $25^{\rm th}$, and $75^{\rm th}$ percentile values of the metallicity gradients for YSs (\gradYS) and SFG (\gradSF) in units of \dexkpc.}
\label{tab:table_stats_gradients}
\begin{tabular}{llr@{\hspace{12pt}}rrr@{\hspace{12pt}}rrr}
\toprule
\toprule
& & & \multicolumn{3}{c}{\gradYS} & \multicolumn{3}{c}{\gradSF} \\
\cmidrule(lr){4-6} \cmidrule(lr){7-9}
Stellar mass bin & Galaxy type & Count & Median & $25^{\rm th}$ p. & $75^{\rm th}$ p. & Median & $25^{\rm th}$ p. & $75^{\rm th}$ p. \\
\midrule
Low-mass & All & 107 & $\sim0$ & -0.026 &  0.022 & $\sim0$ & -0.020 &  0.012 \\
         & NN  &  32 & -0.026  & -0.043 & -0.010 & -0.018  & -0.035 & -0.007 \\
         & NP  &  25 & -0.023  & -0.042 & -0.014 &  0.012  &  0.007 &  0.024 \\
         & PP  &  25 &  0.020  &  0.005 &  0.034 &  0.013  &  0.007 &  0.020 \\
         & PN  &  25 &  0.027  &  0.014 &  0.040 & -0.021  & -0.052 & -0.007 \\
\midrule
High-mass & All & 132 & -0.007 & -0.037 &  0.016 & $\sim0$ & -0.012  &  0.014 \\
          & NN  &  31 & -0.038 & -0.068 & -0.016 & -0.016  & -0.030  & -0.011 \\
          & NP  &  44 & -0.022 & -0.063 & -0.009 &  0.013  &  0.005  &  0.021 \\
          & PP  &  28 &  0.018 &  0.008 &  0.040 &  0.012  &  0.009  &  0.023 \\
          & PN  &  29 &  0.024 &  0.005 &  0.044 & -0.010  & -0.022  & -0.006 \\
\bottomrule
\end{tabular}
\end{table*}

\subsection{SFR distribution}
\label{subsec:sfr}

\begin{figure}
    \centering
    \includegraphics[width=0.9\columnwidth]{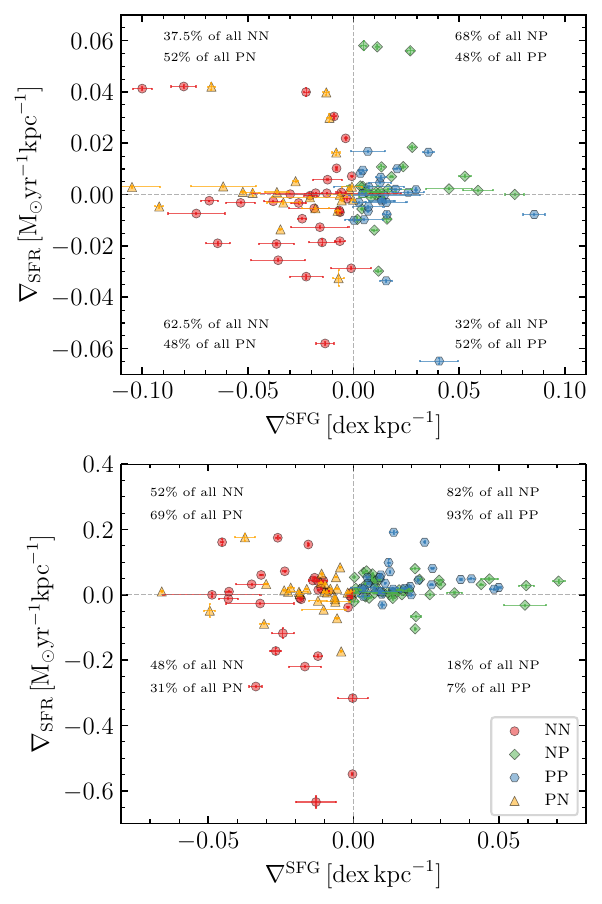}
    \caption{SFR gradient (\gradSFR) as a function of the SFG metallicity gradient (\gradSF) for low-mass (upper panel) and high-mass (lower panel) galaxies. Data points represent galaxies classified by the signs of their young stellar and gas metallicity gradients: NN (red circles), NP (green diamonds), PP (blue hexagons), and PN (orange triangles). The error bars associated with \gradSFR~denote three times the bootstrap errors.}
    \label{fig:grad_sfr}
\end{figure}

The spatial distribution of star-forming regions within a galaxy is key to understanding the radial variation of chemical abundances, as traced by \HII~regions and young stellar populations, which we investigated through the gradient of the SFR: \gradSFR~$[\rm M_{\odot}\ yr^{-1}\ kpc^{-1}]$. A negative \gradSFR~implies a higher level of star formation activity in the central region, potentially driven by recent gas inflows triggered by interactions, mergers, bars, or external accretion. Positive \gradSFR~highlights the inverse situation, with more quiescent central regions compared to the outskirts, which can result from gas depletion due to star-forming events in the past, from heating and expulsion of material by SNe and/or AGN feedback, and via cosmological gas inflows \citep[e.g.][]{sanchez-almeida2014, tapia2025}.

We analysed the \gradSFR~alongside the young stellar and gas metallicity gradients. To do this, we estimated a global \gradSFR~by using the SFG and the youngest stars with ages $<1$ Gyr as tracers of star formation, separately, within the inner $[0-1]$ \Reff~and the outer $[1-2]$ \Reff~regions of each disc. For the stellar-based estimate of the SFR gradient, we selected the age bin as $<1$ Gyr as this provides a reasonable compromise between reducing statistical noise from low-density regions and avoiding over-smoothing of the star formation signal. For each galaxy, we computed the difference between the two estimates: $\nabla_{\rm SFR}^{\rm stars} - \nabla_{\rm SFR}^{\rm gas}$. The median of this difference is $0.007 \pm 0.021\,\rm M_{\odot}\ yr^{-1}\ kpc^{-1}$ for low-mass galaxies and $0.058 \pm 0.330\,\rm M_{\odot}\ yr^{-1}\ kpc^{-1}$ for more massive systems, where the uncertainties correspond to the $1\sigma$ scatter of the distribution. To avoid the impact of a low number of young stellar particles in some systems, we adopted \gradSFR~values derived from the SFG hereafter.

Figure~\ref{fig:grad_sfr} shows \gradSFR~as a function of \gradSF~for low-mass and high-mass galaxies. The different types of galaxies are coloured as in Fig.~\ref{fig:quadrants}. In the low-mass bin, NN galaxies predominantly exhibit negative \gradSFR, with only 37\% showing positive values. In contrast, NP galaxies tend to favour a positive \gradSFR, with 68\% displaying positive slopes. PP and PN systems show a wide range of \gradSFR~values, with a roughly equal divide between positive and negative slopes (48\% and 52\% with positive \gradSFR, respectively). Overall, low-mass NP galaxies tend to have a flatter or positive \gradSFR~compared to the rest of the sample, while NN galaxies display more negative SFR gradients.

For massive galaxies, PP systems have a higher fraction of positive \gradSFR, with $\sim93\%$ of the subsample showing positive values. NP galaxies follow a similar trend, with $\sim82\%$ of them showing positive \gradSFR. Together, PP and NP galaxies predominantly populate the \gradSFR > 0 region, suggesting that massive galaxies with inverted gas gradients tend to avoid negative \gradSFR~and are preferentially located in the upper right quadrant of Fig.~\ref{fig:grad_sfr}. PN galaxies also favour positive \gradSFR, with $\sim69\%$ of the subsample showing positive slopes. NN galaxies exhibit a diversity of SFR gradients, with nearly equal numbers of positive and negative \gradSFR~values.

\subsection{Efficiency of star formation}
\label{subsec:sfe}

\begin{figure}
    \centering
    \includegraphics[width=\columnwidth]{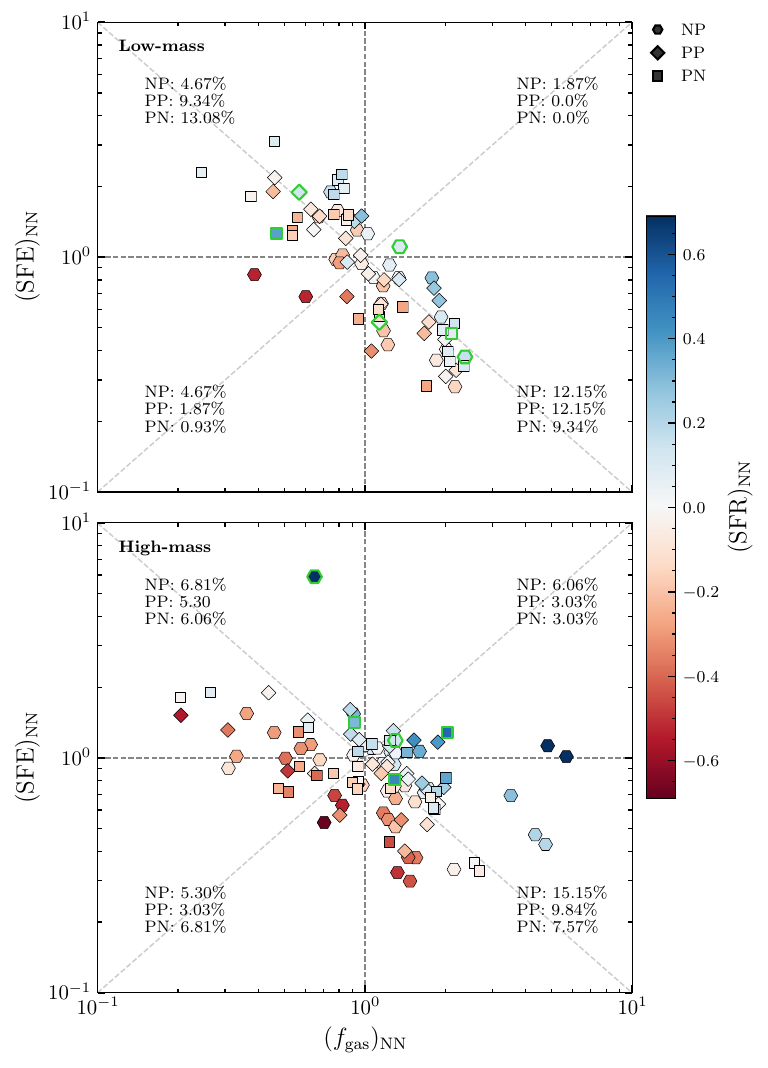}
    \caption{SFE as a function of gas fraction (\fgas) for PP, NP, and PN galaxies in the low-mass (upper panel) and high-mass (lower panel) bins. All parameters are expressed relatively to those of NN galaxies at a given \mstar. The coloured map represents the normalised SFR, while green borders around the data points highlight galaxies that increased their stellar mass by more than $25\%$ within $2.5$ Gyr ($\tau_{25}<2.5$ Gyr). The percentages of each galaxy type within each quadrant are included.}
    \label{fig:fgas_sfe}
\end{figure}

We assessed whether galaxies with specific combinations of young stellar and gas metallicity gradients have different star formation efficiencies (SFEs), where $\mathrm{SFE}\propto\tau_\mathrm{dep}^{-1}$. To do this, we compared the distributions of NP, PP, and PN galaxies relative to NN galaxies in the SFE--\fgas~plane. We took the distribution of SFE and \fgas~of NN galaxies at a given \mstar~as a reference, assuming that NN galaxies show the expected distribution of radial chemical abundances, with both YSs and SFG  having negative gradients.

First, we binned the NN galaxies in stellar mass intervals of 0.25 dex and estimated the median SFE and the median \fgas~in each bin. Then, we normalised the SFE and \fgas~of the other galaxy types to these reference medians according to their stellar mass. This allowed us to determine whether galaxies with different alignments are preferentially located in specific regions of the SFE--\fgas~plane. From our previous analysis (see Fig.~\ref{fig:cdf_all}), we know that NN galaxies tend to have a high SFE. Hence, NN galaxies are efficiently forming stars, though they are not particularly gas rich. 

Figure~\ref{fig:fgas_sfe} displays the relation between the normalised SFE and the normalised \fgas~separately for low-mass and high-mass galaxies. Galaxies are coloured by their normalised SFRs. We note that the SFE and \fgas~determine a clear anti-correlation, as expected, but with different slopes. The slope of this trend differs between the low-mass and high-mass bins, with low-mass galaxies exhibiting a slightly steeper relation. To quantify this, we used the Theil--Sen method \citep{Theil1950,Sen1968} to perform robust linear fits in log–log space within each mass bin. We estimated the respective slope and 95\% confidence intervals (CIs) of the relation using a bootstrap technique, which yields a slope of $-1.033$ and $\rm CI = [-1.189, -0.904]$ for low-mass galaxies and $-0.369$ and $\rm CI= [-0.476, -0.248]$ for high-mass ones.

In general, the most populated quadrants correspond to galaxies with high SFEs and low gas reservoirs, and vice versa, with respect to NN galaxies. At a given \fgas, galaxies with an excess of star formation activity (bluer colours) have higher SFE, and conversely, galaxies with lower star formation activity with respect to NN systems have low SFEs. Additionally, we highlight galaxies that experienced an increase in their stellar mass by more than $25\%$ within the last $2.5$ Gyr of their evolution ($\tau_{25}<2.5$ Gyr). 

As can be seen in the upper panel of Fig.~\ref{fig:fgas_sfe}, NP and PP galaxies are more frequent in the low-SFE and high-\fgas~quadrant, while PN galaxies are more frequent in high-SFE and low-\fgas~quadrant. If we consider the latter as galaxies that are in the process of recovering their negative metallicity gradients, then this suggests that they are doing so by increasing the efficiency of star formation with respect to NN types. The low-mass galaxies with recent violent events (i.e. $\tau_{25}<2.5$ Gyr) tend to be actively forming stars, as expected. There are also a few galaxies with recent violent events, high SFEs, and low \fgas, which coexist with galaxies that have a more quiescent recent history at the same gas richness. As mentioned above, low-mass galaxies follow the 1:1 anti-correlation, which suggests that they follow a similar trend to the NN galaxies overall in the way they regulate their star formation activity.

In the lower panel of Fig.~\ref{fig:fgas_sfe}, high-mass galaxies are displayed. As mentioned above, the slope of this relation is much shallower, indicating that galaxies with an excess of gas are more efficient than NN galaxies. A significant fraction of them are NP systems, which are thought to have recently experienced a strong starburst or gas inflow that inverted the gas metallicity gradients. Galaxies with recent violent events have been highlighted. Galaxies with lower \fgas~than NN tend to be even less efficient, with lower star formation activity.

\subsection{AGN feedback}
\label{subsec:agn_activity}

\begin{figure}
    \centering
    \includegraphics[width=0.9\columnwidth]{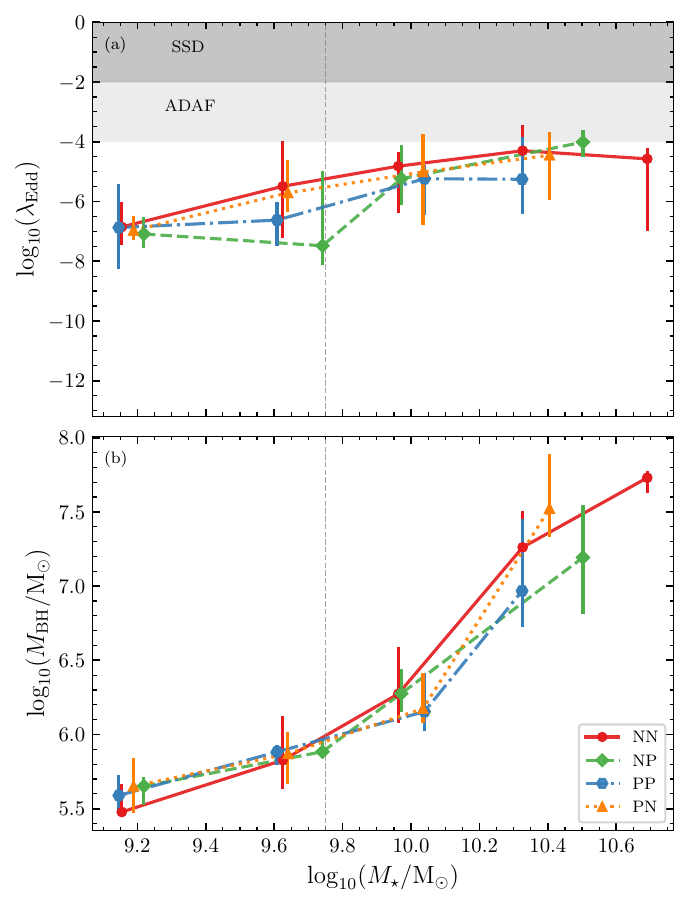}
    \caption{Upper panel: Eddington parameter ($\Ledd$) as a function of galaxy stellar mass (\mstar). The dark grey shaded region indicates the regime of radiatively efficient (SSD) accretion, while the light grey region marks the radiatively inefficient (ADAF) regime. Lower panel: BH mass as a function of \mstar~(the $\mbh$--\mstar relation). Median trends for each galaxy subsample are shown, where the median values were estimated by dividing the stellar mass into five bins. Error bars represent 95\% CIs for the medians estimated via bootstrap resampling. The dashed vertical lines in both panels indicate the stellar mass threshold separating the low-mass and high-mass subsamples.}
    \label{fig:figure_AGN}
\end{figure}

We analysed the AGN activity in our galaxy sample to study its role in shaping the young stellar and gas metallicity gradients. For this purpose, we estimated the Eddington ratio\footnote{Ratio between the luminosity of a BH and its Eddington luminosity, which is the maximum theoretical luminosity that it can reach while maintaining hydrostatic equilibrium. It can also be expressed as the ratio of BH accretion rate and the Eddington limit \citep{Rosas-Guevara2016}.}, $\Ledd$, using the central BH mass, $\mbh$, and the BH accretion rate, $\mdotbh$, for each galaxy in our sample as 
\begin{equation}
    \Ledd \equiv \frac{\mdotbh}{\mdotedd} = \frac{\epsilon_{\rm r} c \sigma_{\rm{T}} \mdotbh}{4 \pi G m_{\rm{p}} \mbh},
\end{equation}
where $\mdotedd$ is the Eddington accretion rate\footnote{Maximum rate of accretion of material onto a compact object before the radiation pressure from the infalling material balances the object's gravitational pull.}, $\epsilon_{\rm r}$ is the radiative efficiency of the BH set to 0.1, $\sigma_{\rm T}$ is the Thompson scattering cross-section, and $m_{\rm p}$ is the proton mass.

Following \citet{Rosas-Guevara2016}, we considered two primary regimes for active accretion onto supermassive BHs based on their $\Ledd$. For $\Ledd > 10^{-2}$, accretion is considered radiatively efficient, with the nuclear disc described by the standard Shakura–Sunyaev thin-disc (SSD) model \citep{shakura1973}. In this regime, cooling is efficient, and the disc emits strongly in X-rays, which is typical of luminous AGNs and quasars accreting near the Eddington limit. For $\Ledd=[10^{-4},10^{-2}]$, accretion is assumed to proceed via a radiatively inefficient mode, which is characterised by geometrically thick, hot flows with long cooling times, called advection-dominated accretion flows (ADAFs; \citealt{narayan1994}). These flows advect much of their thermal energy into the BH rather than radiating it away and are associated with low-to-intermediate-luminosity AGNs. Galaxies with $\Ledd < 10^{-4}$ in the sample are considered inactive. \citet{rosasguevara2019} reported a large variability in the activity of supermassive BHs. Hence, we took these trends as indicative only.

In Fig.~\ref{fig:figure_AGN} we show \mstar~as a function of $\Ledd$ and $\mbh$ for galaxies grouped according to their combinations of metallicity gradients. We note that, on average, the level of activity (i.e. $\Ledd$) increases with increasing stellar mass. We note that our sample does not contain PP-type galaxies with \mstar $>10^{10.5}$ \Msun.

Regarding the upper panel of Fig.~\ref{fig:figure_AGN}, we find that PP galaxies contribute less within the ADAF regime ($\sim17\%$), while NN, NP, and PN galaxies contribute similarly within this regime ($\sim29\%$, $\sim27\%$, and $\sim27\%$, respectively). In the radiatively efficient SSD regime, only four massive galaxies are present, three of which are NP galaxies. However, we note that significant error bars prevented us from establishing clear trends.

Regarding $\mbh$ in the lower panel of Fig.~\ref{fig:figure_AGN}, there is a trend for massive galaxies with negative SFG gradients to have slightly more massive central BH compared to other galaxy types at a given \mstar. Overall, due to the uncertainties, we cannot draw a robust conclusion on the role of AGNs beyond reporting a more significant impact on NP galaxies and the fact that PP galaxies seem to be potentially less affected by AGN feedback in the recent past, at least.

\section{Conclusions}
\label{sec:conclusion}

We studied the radial metallicity gradients of star-forming regions and stars younger than $2$ Gyr in galaxies selected from the higher resolution run of the \eagle~project. We found a diversity of behaviour with galaxies having aligned or misaligned metallicity gradients in YSs and SFG. We acknowledge that our results might depend on the adopted sub-grid physics, and hence they provide a way of testing it.
Our results can be summarised as follows:

\begin{itemize}

    \item NN galaxies exhibit negative metallicity gradients in both YSs and SFG, indicating that they have been actively forming stars for the last $2$ Gyr while maintaining these profiles, be they steep or shallow. They have a high SFE at all stellar masses, driven by a combination of high SFR and low \fgas. These galaxies tend to have a significant bulge. Such NN profiles are characteristic of systems that grow globally from the inside out.

    \item NP galaxies exhibit inverted metallicity gradients in their star-forming regions and show positive \gradSFR. On average, they also have the longer gas-depletion times. The persistence of negative metallicity gradients in YSs suggests that the SFE has remained low over the past $2$ Gyr --as reflected by the long depletion times-- or that the inversion of the SFG gradients occurred more recently.

    \item PP galaxies exhibit inverted metallicity gradients in both components. They tend to be slightly less massive, which supports the hypothesis that SN feedback is the dominant mechanism driving the ejection of material and inverting both gradients. This is further evidenced by the fact that these galaxies host smaller BHs and exhibit lower Eddington ratios than the rest of the sample at a given stellar mass. This implies that AGN feedback is unlikely to be the primary driver of their evolution. Some low-mass PP galaxies exhibit negative \gradSFR, suggesting that a significant gas component is available for star formation. However, the positive \gradYS~indicates that this process began only recently.

    \item PN galaxies have inverted metallicity gradients in their YSs, implying that they experienced intense star formation activity in the outer regions over the past $2$ Gyr. Their positive \gradSFR~indicates ongoing star formation in their outskirts, yet the SFG itself displays negative metallicity gradients. In the case of low-mass galaxies, one explanation is that hot, turbulent gas has begun to cool and become available for star formation. In high-mass galaxies, this behaviour may result from metal-rich inflows triggered by interactions that occurred more than $2$ Gyr ago, which subsequently fuelled central star formation as the gas reached the inner regions. This points to the occurrence of a violent event more than $2$ Gyr ago that increased the gas fraction and reversed the stellar metallicity gradient, followed by a more recent recovery of the negative metallicity gradient in the gas phase.
    
\end{itemize}

\citetalias{tissera2022} estimated a median timescale of $\sim2$ Gyr with $25^{\rm{th}}$ and $75^{\rm{th}}$ percentiles equal to 1.4 Gyr and $3$ Gyr, respectively, for the evolution of the metallicity gradient, which suggests that galaxies could undergo cycles of steepening, flattening, and inversion driven by violent events in their history. The characteristics of this cycle depend on properties such as the stellar mass, the gas fraction, and the orbital parameters of the infalling material. Within this cycle, NP will be the first stage, and galaxies could return to being NN types if the system recovers the negative slopes for the gas components quickly (i.e. the starbursts are not strong enough for the potential well of the systems to blow material). Otherwise, YSs can determine a positive gradient, and this implies a stronger impact on the systems with longer recovery timescale and/or higher SFEs. PP galaxies have inverted gradients in both components, which implies the strongest impact of feedback in systems that tend to be less massive. PN galaxies could be in a stage of recovery after a strong feedback event and inversion of the metallicity gradient with gas cooling down in the central region. Galaxies with misaligned or inverted metallicity gradients in YSs and SFG show either recent strong accretion or mergers with high \fgas~or SFE compared to galaxies with aligned negative gradients.

Another important source of feedback includes AGN events, but we cannot draw a definitive conclusion regarding their role, beyond noting their more significant current impact on massive NP galaxies and the indication that PP galaxies have likely been less affected by AGN feedback in the recent past. We plan to address this aspect in a future work by using another simulated galaxy sample.

In this paper, we present a novel approach to dating the impact of recent violent events. The comparison between stellar and gas metallicity and metallicity gradients is relatively recent from an observational \citep{molla2020, sextl2024, Zinchenko&Vilchez2024} and numerical point of view \citep{lian2023}. According to our analysis, the study of young stellar and gas metallicity slopes in tandem stores information on both the recent star formation history and mixing processes, as well as the sub-grid physics adopted in simulations of galaxy formation.

\begin{acknowledgements}
PBT acknowledges partial funding by Fondecyt-ANID 1240465/2024, N\'ucleo Milenio ERIS NCN2021\_017, and ANID Basal Project FB210003. We acknowledge support from the European Research Executive Agency HORIZON-MSCA-2021-SE-01 Research and Innovation programme under the Marie Skłodowska-Curie grant agreement number 101086388 (LACEGAL). 
ES acknowledges funding by Fondecyt-ANID Postdoctoral 2024 Project N°3240644 and thanks the N\'ucleo Milenio ERIS.
This project used the Ladgerda Cluster (Fondecyt 1200703/2020 hosted at the Institute of Astrophysics, Chile).
\end{acknowledgements}



\bibliographystyle{aa}
\bibliography{bibliography.bib}

\let\oldbibitem\bibitem


\newpage
\begin{appendix}

\section{Normalised metallicity gradients}

\begin{figure}
        \includegraphics[width=\columnwidth]{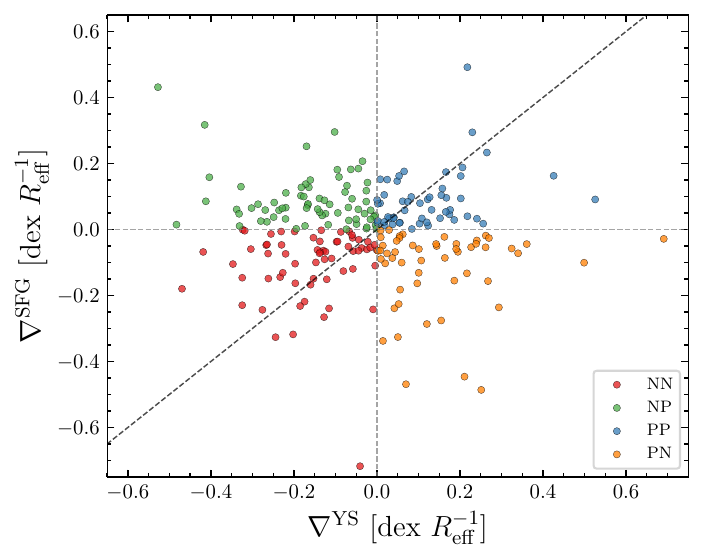}
    \caption{Normalised metallicity gradients of SFG, \gradSF~(\dexReff), as a function of normalised metallicity gradients of YSs, \gradYS~(\dexReff), in our \eagle~sample.}
    \label{fig:quadrantsnorm}
\end{figure}

Similar to Fig.~\ref{fig:quadrants}, we show \gradSF~as a function of \gradYS~in Fig.~\ref{fig:quadrantsnorm}, with both gradients normalised by each galaxy's half-mass radius, \Reff, and thus expressed in units of \dexReff, following previous works \citep[e.g.][]{SanchezMen2018}. As the figure shows, there is no clear correlation between the two gradients, in agreement with the above results. We note that since we only focus on the sign of the metallicity gradients, this normalisation does not affect our conclusions.

\label{lastpage}
\end{appendix}
\end{document}